\documentclass[12pt]{article}

    \newcommand\beq{\begin{eqnarray}}
    \newcommand\eeq{\end{eqnarray}}

\bibliography{unsrt}
\begin{document}
\vspace*{3cm}

\begin{center}

{\Large \bf Neutrino Properties, Cosmology}\\
\vspace{1cm}
{\large Ernest M. Henley \\
Physics Department, Box 351560\\
University of Washington, Seattle, WA 98195}
\\[0.2cm]
\end{center}

\vspace{1cm}

\begin{abstract}
This is a review of our present knowledge of neutrino properties and what remains to be determined. It  is followed by a description of double beta decay and very high energy neutrinos. Finally, there is a description of leptogenesis. In this talk I will neglect sterile neutrinos, and thus the LSND problem. 
\end{abstract}
\vspace {1 cm}

\section{Neutrino Properties}

We now know that neutrinos have masses and that the flavor eigenstates $(\nu_e, \nu_\mu, \nu_\tau)$  are not the mass eigenstates ($\nu_1, \nu_2, \nu_3$). However, we still do not know the neutrino masses, but only the squares of their mass differences.  Thus, we still have to learn whether the order of the mass eigenstates is normal (1,2,3),  inverted (3,1,2), or quasi-degenerate; see Fig. 1. SNO \cite {SNO}and KamLAND \cite{Kam} have shown us that $m_2>m_1$.

\begin{figure}
\unitlength1cm 
\begin{picture}(8,8)(-2,-4)
  \includegraphics{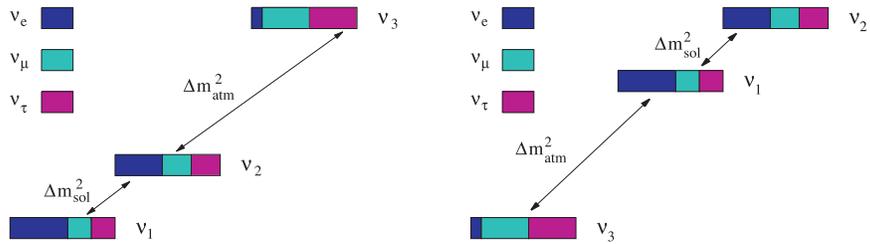}
\end{picture}
\label{fig:1}
\caption{Normal and inverted neutrino masses; from E.K. Akhmedov, hep-ph/0610064 }
\end{figure}

From the solar
neutrino experiments, we also know that $\theta_{13}$ is small; CHOOZ has set a limit , $\sin^2  \theta_{13} \leq .006^{+.03}_{-.006}$ \cite{3}. It is thus a reasonable 
approximation to use two flavor eigenstate mixing
\beq
\mid\nu_e> &=& \cos\theta_s \mid\nu_1> + \sin \theta_s \mid \nu_2 >\;, \nonumber\\
\mid \nu_\mu> &=& -\sin\theta_s\mid \nu_1> + \cos\theta_s \mid \nu_2 > \; ,\\
{\cal P}_{\mu_ e}&=& \sin^2 2\theta_s \sin^2(\frac{\bigtriangleup m_{12}^2}{4E}L)\;,
\eeq
where $m_{12}= m_2 -m_1$, E is the energy of the neutrinos, L is the distance they have traveled, and $\theta_s$ is the mixing angle for solar neutrinos. ${\cal P}_{ab}$ indicates the probability of a transition from $a$ to $b$. 

In matter, the MSW effect \cite {MSW} enters because the electron neutrino interacts with electrons not only via weak neutral currents, but also charged currents. The charged current interaction gives rise to an effective potential, $V=\sqrt{2} G_F \rho_e$, where $G_F$ is the Fermi coupling constant and $\rho_e$ is the electron density. In the presence of the MSW effect, the equation of motion for the two neutrino flavors is

\begin{equation}
i \frac{d}{dt} \left |
 \begin{array}{l}
\nu_e\\
\nu_\mu
\end{array} \right) 
=
  \left( 
\begin{array}{cl}
\cos 2\theta_s + V \frac{4E}{\bigtriangleup m_{12}^2}
& \sin 2\theta_s\\
-\sin 2\theta_s& \cos  2\theta_s
\end{array} \right)
\left |
\begin {array}{l}
 \nu_e\\
     \nu_\mu 
\end{array} \right ) \; .
\end{equation}

With the MSW effect, the angle $\theta$ effectively becomes $\theta_m$, where the $ m$ stands for matter,
\begin{equation} 
\sin^2 2\theta_{s,m}= \frac{\sin^2 2\theta_s}{(\cos 2\theta_s -\sqrt{2} G_F \rho_e\frac{2E}{\bigtriangleup m_{12}^2})^2 + \sin^2 2 \theta_s} \; .
\end{equation}

\begin{figure}
\unitlength1cm 
\begin{picture}(8,8)(-2,0)
  \includegraphics{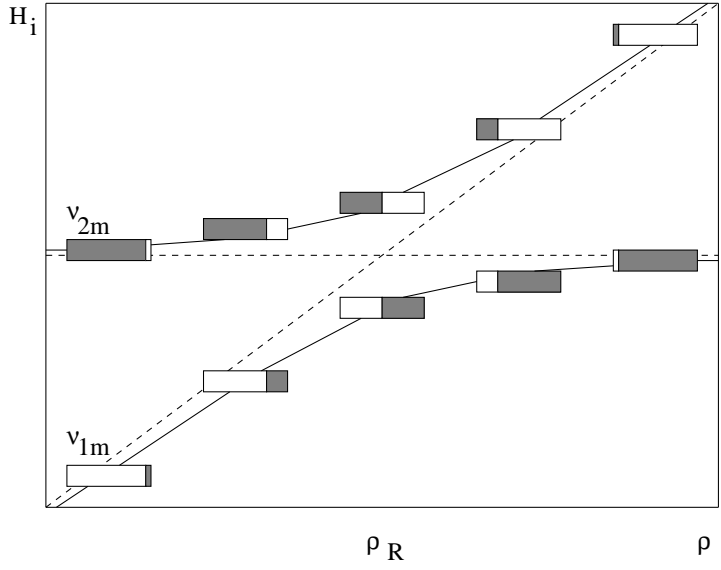}
\end{picture}
\label{fig:2}
\caption{ Adiabatic neutrino flavor conversion; from E.K. Akhmedov, hep-p/0610064 }
\end{figure}

As the density of electrons increases, the electron neutrino ($\sim \nu_1$) becomes more massive than the muon neutrino($\sim \nu_2$); see Fig. 2. This may occur in the center of the sun for $^8B$ neutrinos. Also, as the energy of the electron neutrinos increases, the MSW effect rises in importance. It is likely to be important in the $^8B$ data of the solar neutrinos because for vacuum oscillations 
\begin{equation}
{\cal P}_{ee} \approx 1 - \frac{1}{2} \sin^2 2 \theta_s \geq 1/2\; ,
\end{equation} 
and the data from SNO suggests that ${\cal {P}}_{ee} \leq 1/2$\cite{5}.
A full analysis of the SNO and Kamiokande data give 
\beq
|  \bigtriangleup m_{12}^2 | &=& (7.9 \pm .5) 10^{-5} eV^2\; ,\\
  \sin^2\theta_{12} &=& 0.30 \pm .08 \; (\theta_{12} \approx 36.9^o)\; .
\eeq
From the atmospheric neutrino data, one finds $\nu_\mu \rightarrow \nu_\tau$ 
and 
\beq
| \bigtriangleup m_{23}^2|  &=& (2.6 \pm 0.5) 10^{-3} eV^2, \\
\sin^2 \theta_{23}& =& 0.52 \pm 0.2 \; (2\theta_{23} \approx 90^o)\; .
\eeq
If $2\theta_{23}$ is exactly equal to $ 90^o$, we would want to know why that is so.

Experiments have gotten to a sufficiently high precision that we now require the $ 3 \times 3$ neutrino mixing matrix

\begin{equation}
\mid \nu_\alpha^{fl} > = \sum_{i=1}^3 U_{\alpha i}^* \mid \nu_i^{mass} >,
\end{equation}\\
\begin{equation}
U = 			
\left( 
\begin{array}{lll}
c_{12}c_{13}   & s_{12}c_{13}  & s_{13}e^{-i\delta}\\
-s_{12}c_{23}-c_{12}s_{23}s_{13}e^{i \delta}  &  c_{12}c_{23}-s_{12}s_{23}s_{13} e^{i\delta}  & s_{23} c_{13}\\
s_{12}s_{23} -c_{12}c_{23}s_{13}e^{i\delta} & -c_{12}s_{23} -s_{12}c_{23}s_{13} e^{i\delta}  &   c_{23}c_{13}\\
\end{array}\right) \\
\times diag(1, e^{i \alpha_2}, e^{i \alpha_3})\; ,\\
\end{equation}
where $s_{ij}$ stands for $\sin\theta_{ij}$ and $c_{ij}$ for $\cos \theta_{ij}$.

It is important to measure $s_{13}$ and obtain as accurate a value as possible for it. All CP non-conservation tests depend on it. The proposed experiments to obtain more than an upper limit  require high precision, as in Chooz, Double Chooz, T2KK, MINOS and OPERA \cite{7}. In the $3 \times 3$ case, we have, for example
\begin{equation}
{\cal P}_{ee} \approx  1- \sin^2 2 \theta _{13}\sin^2 \frac{\bigtriangleup m_{31}^2 }{4 E}L \; .
\end{equation}
The hope is that these measurements can reach $\sin^2 2 \theta_{13} =10^{-3}$.
There is no reason why $\theta_{13}$ should be zero.  Once it is known, it becomes possible to measure CP and T violation for neutrinos. For instance,
\begin{equation}
{\cal P}(\nu_a \rightarrow \nu_b) - {\cal P} (\nu_b \rightarrow \nu_a)
\end {equation} 
tests time reversal invariance, but care is required that matter effects do not give a false signal. Similarly
\begin{equation}
{\cal P} (\nu_a \rightarrow \nu_b) - {\cal P}(\bar{\nu}_a \rightarrow \bar{\nu}_b)
\end{equation}
tests CP conservation.

In addition to $\theta_{13}$, we still need to learn the hierarchy of neutrino masses \cite{9}.  In ascending order, are the masses 1-2-3- (normal ordering), 3-1-2 (inverted ordering) or are the masses quasi-degenerate?  Here, long baseline experiments with an interference of matter and vacuum oscillations can help us determine the hierarchial sequence of masses.  Thus, for example
\beq 
{\cal P}(\nu_e \leftrightarrow \nu_\mu) &=& \sin^2 \theta_{23} \sin^2 2\theta_{13}^m \sin^2 \frac{\bigtriangleup m_m^2}{4E} L\;,\\
\sin^2 2 \theta_{13}^m &=& \frac{\sin^2 2\theta_{13}}{(\cos 2\theta_{13} - \frac{2 \sqrt{2} G_F \rho_e E_\nu}{\bigtriangleup m_{13}^2})^2 + \sin^2 2\theta_{13}}
\eeq

Ultimately, we would, of course, like to know the actual masses of the neutrinos. From the beta decay of $^3 H$, we know that $m_{\nu_e} \leq 2.2 eV$ \cite{Bonn}. The KATRIN collaboration \cite{10} expects to lower this limit to 0.35 eV. On the other hand,WMAP together with the Galactic Red Shift Survey gives $\sum_{i=1}^3 m_i \leq 0.70 eV$ \cite{11}. 

\section{Neutrinoless Double Beta Decay}

Neutrinoless double beta decay violates lepton number conservation; indeed $\bigtriangleup L = 2$, and the process requires massive Majorana neutrinos \cite {12}. The rate of the reaction depends on $m_{ee}^2$,
\begin{equation}
m_{ee}^2 = |\sum_{i=1}^3 U_{ei}^2 m_i  |^2= (| U_{e1}|^2 m_1 + |
U_{e2} |^2 e^{i\alpha_2} m_2 + | U_{e3}|^2 m_3 e^{i \alpha_3})^2 \; .
\end{equation}
This should be contrasted to ordinary beta decay, which is sensitive to $m_e^2$
\begin{equation}
m_e^2= m_1^2 \mid U_{e1} \mid^2 + m_2^2 \mid U_{e2} \mid^2 + m_3^2 \mid U_{e3}\mid^2\;.
\end{equation} 
The most precise limit has been established by the Moscow-Heidelberg group for $^{76} Ge$, with $T_{1/2} \geq 1.9 \times 10^{25} y$ \cite {13}, corresponding to 
$m_{\nu(e)} \leq 0.35 eV$ if the matrix element is interpreted liberally. However, 
a subgroup of the Moscow-Heidelberg collaboration measured a neutrinoless double beta decay rate for $^{76}Ge$ \cite{14} with a half life of $(0.7-4.2) \times  10^{25} y$, which would mean $m_{ee} = (0.2 -  0.6) eV$. In view of the WMAP and Galactic Red Shift Survey results, this would imply that the neutrino masses are quasi-degenerate.
Clearly, this experiment is being repeated by other groups: COBRA, CUORICINO, NEMOS in Gran Sasso and elsewhere. Other nuclei are also being used, e.g.,  $^{130}Te, ^{116}Cd, ^{100} Mo$ \cite {15}.

\section{Cosmic Rays, Superhigh Energy Neutrinos}

The spectrum of cosmic rays is shown in Fig.3 \cite {16}. There is a ``knee'' and an ``ankle'' in the spectrum,  but the flux follows a power law; it is proportional to $E^{-n}$, with $n \sim 2$. For energies above the ankle, the cosmic rays are expected to be primarily protons and not nuclei, as at lower energies. Where there are protons, you can expect to find neutrinos, since $ p + \gamma \rightarrow \pi +...\rightarrow \nu $, where the photons are cosmic ray background.  

\begin{figure}
\unitlength1cm 
\begin{picture}(8,12)(-4,0)
  \includegraphics{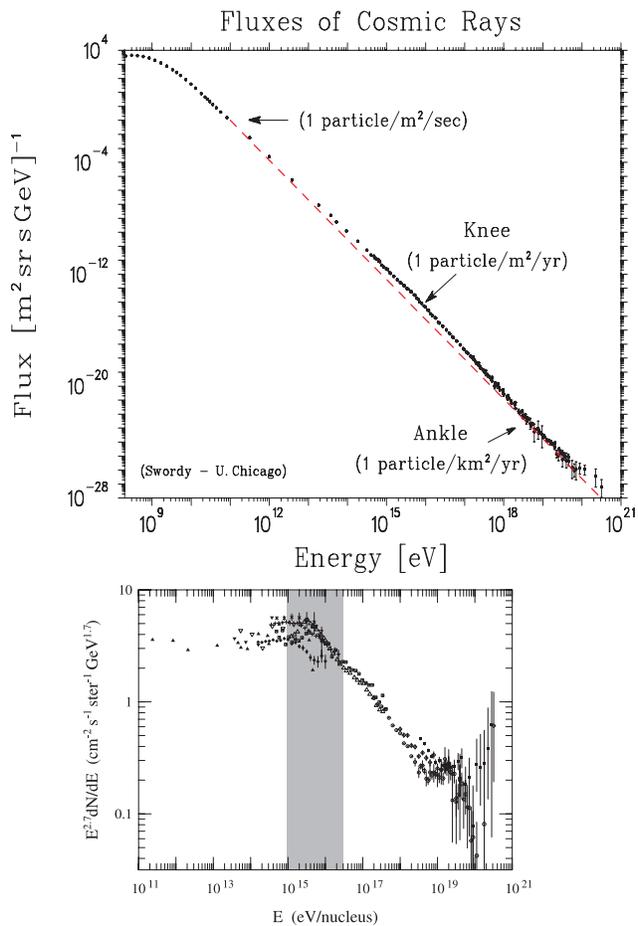}
\end{picture}
\label{fig:3}
\caption{The cosmic ray spectrum; courtesy of S.P. Swordy and from 
T. K. Gaisser and T. Stanev, Nucl. Phys. {\bf{A777}}, 98 (2006) }
\end{figure}

Unlike protons, neutrinos can penetrate deep into the atmosphere, so that we expect more lateral showers for neutrinos. Large arrays are being built to detect high energy neutrinos. Examples are AGASA and HIRES \cite{17}. For protons , there is a high energy cut-off , the GZK limit \cite {GZK} at about $4 \times 10^{19} eV$ because protons can annihilate with background photons, e.g., $2E E_\gamma = E_{cm}^2$ with $E_{cm} \sim m_\bigtriangleup = 1232 MeV$
  at the GZK energy.  Thus, pions will be produced readily  at and above this energy.

An interesting questions is the origination of cosmic rays with energies $\geq 10^{17} eV$ . The Fermi mechanism stops working because it becomes impossible to confine the charged particles in the  magnetic field. This suggests that very high energy cosmic rays are extra-galactic. The observed near isotropy of the cosmic rays also suggests a cosmological origin. Possible sources are supernovae, pulsars, gamma-ray bursts, and black holes \cite{18}. Are the highest energy cosmic rays neutrinos? This is not ruled out \cite{16}. AGASA observed neutrinos above the GZK limit,(see Fig. 3), but these results
are not yet confirmed. Is the energy correct? There are two problems for experimentalists to set the energy scale. The first problem is the small flux, so that errors are large, and the second one is the dependence of the flux on the neutrino-proton cross section. The cross section is sensitive to the  parton distribution function in the proton, particularly that of gluons at very small values of Feynman $x$ in the color dipole model \cite{19}. Does the distribution keep rising as $x$  decreases  beyond $x = 10^{-5}$? If so, then the cross section can get to be of the order of hadronic ones  This 
uncertainty needs to be removed. 

One possibility for producing superhigh energy neutrinos is the so-called Z-burst mechanism \cite{20}, where the cm energy of neutrinos scattering on background neutrinos is equal to the mass of the Z-boson,
\begin {equation}
E_\nu = \frac{M_Z^2}{2 m_\nu}\;,
\end {equation}
for enhanced production. If the neutrino mass is taken as $10^{-2} eV$, then $ E_\nu \approx 10^{23} - 10^{24} eV$.

Some open questions thus involve the sources of very high energy neutrinos, neutrinos observed above the GZK limit, and the interaction cross section of such superhigh energy neutrinos with the atmosphere. 

\section {Baryogenesis via Leptogenesis}

Leptogenesis could be at the root of the excess of baryons over antibaryons in the early universe. This proposal would get a large boost if the neutrinos are 
Majorana rather than Dirac particles. This is because for Majorana neutrinos there is a plausible explanation for the small neutrino masses, dubbed the see-saw model \cite{21}. By now, several variants have been proposed, but here, we stick to the simplest one. The model proposes the existence of right-handed heavy neutrinos, with a mass of the order of $10^{12} -10^{16} GeV$, where the latter mass is of the order of the GUT scale. The neutrino masses are then proposed to be of the order of
\begin{equation}
m_\nu  \sim \frac{v^2}{M_R}   \sim \frac{10^4 GeV^2}{10^{15}GeV} \approx
10^{-11} GeV = 10^{-2} eV \; ,
\end{equation}
where $v$ is a Dirac mass of the order of the $Z^0$ or Higgs boson masses, i.e. 100  GeV.
The minimal see-saw model has two right-handed neutrinos; in that case one of the normal neutrinos would have zero mass. The right-handed neutrinos decay, and we assume that the lightest of them remains in thermal equilibrium when the more massive one decays. However, when $T< M_{R1}$, the lightest right-handed neutrino, (R1), decays out of equilibrium.Thus, we obtain thermal leptogenesis \cite{22}. The decay of R1 is to both $\ell H $ and $\bar \ell H^\dagger$ with different decay rates. Here $\ell$ is a charged lepton and H is a charged Higgs. This difference gives rise to a CP-violating asymmetry due to the interference of the tree level and one loop Feynman diagrams. The Sakharov conditions are thus met ,
\begin{equation}
Y_L = \frac{n_L - \bar{n}_L}{s} \neq 0;.
\end{equation}
Here Y is the lepton-antilepton asymmetry of the universe, $n^L- \bar{n}_L$ is
the lepton number asymmetry , and  s is the entropy of the universe per unit volume. The asymmetry can be transmitted to the baryons to give a $ (B - \bar{B})$asymmetry via the sphalaron 
process \cite{23}, which is a thermal fluctuation over the barrier. The process maintains $B-L$, where B is the baryon and L the lepton number and we have 
\begin{equation}
Y_B = \frac{28}{79} Y_L  \approx 0.35 Y_L \; .
\end{equation}
It is quite possible to meet the known experimental condition, $ \frac {B - \bar{B}}{n_\gamma} \sim 6 \times 10^{-10}$ \cite {22}, even with a reasonable washout factor of the CP violation. 
\section{Summary}

Neutrinos began as a suggestion by Pauli to save energy conservation in beta decay. It took 30 years to discover them. They have now taken center-stage. They are central to our understanding of the universe and its development. We have learned a lot about them in the past decade, but much more information remains to be found. They will continue to play a central role for quite some time. 

I thank Alajandro Garcia for help with the figures. My work is supported, in part, by the Department of Energy.


\begin{thebibliography}{99}
\bibitem{SNO} Q.R. Ahmad et al.(SNO Collab), Phys. Rev. Lett. {\bf{87}}, 071301 (2001);  {\bf{89}}, 011301 (2002); {\bf{89}}, 011302 (2002)

\bibitem {Kam} K. Eguchi et al., Phys. Rev. Lett. {\bf{90}}, 021802 (2003); S. Fukuda et al. , Phys. Rev Lett. {\bf{86}}, 5651 (2001), Phys. Lett. {\bf{B539}}, 179 (2002)

\bibitem{3} M. Apollonio et al, Phys. Lett. {\bf{B420}}, 397 (1998), {\bf {B466}}, 415 (1999)

\bibitem{MSW} L. Wolfenstein, Phys. Rev {\bf{D17}}, 2369 (1978); S.P. Mikheev and A.Y. Smirnov, Yad. Fiz. {\bf{42}}, 1441 (1985)[tr. Sov. J. Nucl. Phys. {\bf{42}}, 913 (1985)

\bibitem{5} See, e.g., D.P. Roy, hep-ph/0409336

\bibitem {7} see e.g.,F. Ardellier et al., hep-ex/0405032;  H. Minakata et al., Phys. Rev. {\bf{D68}}, 033017 (2003), K. Anderson, hep-ex/0402041. 
  
\bibitem{9} H. Minakata, H. Nunokawa, S.J. Parker, and R.Z. Fundal, Phys. Rev {\bf{74}}, 053008 (2006)

\bibitem{Bonn} J. Bonn et al., Progr. Part. Nucl. Phys. {\bf{48}}, 133 (2002); V.M. Lobashev et al., Nucl. Phys. Proce. Suppl. {\bf{91}}, 280 (2001)

\bibitem {10} A. Osipowicz et al. (KATRIN Collab.) hep-ex/0109033

\bibitem {11} O. Elgaroy et al.  Phys. Rev. Lett.  {\bf{89}}, 061301 (2002); J.R. Kristiansen , O. Elgaroy, and H.K. Eriksen hep-ph/0608017

\bibitem {12} S.M. Bilenky hep-ph/0509098; K. Zuber, hep-ex/0610007

\bibitem {13} H.V. Klapdor-Kleingrothaus et al., Eur. Phys. J. {\bf{A12}}, 147 (2001)

\bibitem{14} H.V. Klapdor-Kleingrothaus et al., Mod. Phys. Lett. {\bf{A17}}, 1475 (2002); Phys. Lett. {\bf{B586}}, 198 (2004)

\bibitem {15} see e.g., K. Zuber, Phys. Lett {\bf{B519}}, 1 (2001), nucl-ex/0610007

\bibitem {16} M Ahlers, A.Ringwald,and H. Tu, astro-ph,/0506698; T.K. Gaisser and T. Stanev, Nucl. Phys. {\bf{A777}}, 98 (2006); D. Seckel and T. Stanev, Phys. Rev. Lett. {\bf{95}}, 141101 (2005)

\bibitem {17} B.M. Connolly et al., Phys. Rev. {\bf{D74}}, 043001 (2005)

\bibitem{GZK} K. Greisen, Phys. Rev. Lett. {\bf{16}}, 748 (1966);  G. T.
Zatsepin and V.A. Kuzmin JETP Lett. {\bf{4}}, 78 (1966)

 \bibitem{18} D.F. Torres and L.A. Anchordoqui, astro-ph/0402371

\bibitem {19} E.M. Henley and J. Jalilian-Marian, Phys. Rev. {\bf{D73}}, 094004 (2006)

\bibitem{20} See e.g., Z. Fodor, S.D. Katz, and A. Ringwald, Phys. Rev. Lett. {\bf{88}}, 171101 (2002); S. Hannestad, New J. Phys. {\bf {6}}, 108 (2004)

\bibitem{21} See e.g. R.N. Mohapatra and G. Senjanovic, Phys. Rev. Lett. {\bf{44}}, 912 (1980); H. Tsujimoto, hep-ph/0501023

\bibitem{22} W. Buchm\"{u}ller, R.D. Peccei, and T. Yanagida, hep-ph/0502169; W-l Guo, Z-z Xing, and S. Zhou, Int. J. Mod. Phys. E ( to appear)

\bibitem {23} G. 't Hooft, Phys. Rev. Lett. {\bf{37}}, 8, (1976); J.S. Harvey and M.S. Turner, Phys. Rev. {\bf{D42}}, 3344 (1990)

\end{thebibliography}
\end{document}